\documentclass[10pt]{article}
\usepackage{epsfig}

\begin{document}

\title{\bf Review of results from SND detector}
\author{ M.N.Achasov\thanks{e-mail: achasov@inp.nsk.su}, V.M.Aulchenko,
         K.I.Beloborodov, A.V.Berdyugin, \\
         A.G.Bogdanchikov, A.V.Bozhenok, A.D.Bukin, D.A.Bukin, \\
         S.V.Burdin, T.V.Dimova, A.A.Drozdetski, V.P.Druzhinin, \\
         D.I.Ganushin, V.B.Golubev, V.N.Ivanchenko, P.M.Ivanov, \\
         A.A.Korol, S.V.Koshuba, I.N.Nesterenko, E.V.Pakhtusova, \\
         A.A.Polunin, A.A.Salnikov, S.I.Serednyakov, V.V.Shary, \\
	 Yu.M.Shatunov, V.A.Sidorov, Z.K.Silagadze, A.N.Skrinsky, \\
	 A.G.Skripkin, Yu.V.Usov, A.V.Vasiljev \\
         \it \small
         Budker Institute of Nuclear Physics  \\
         \it \small
         Siberian Branch of the Russian Academy of Sciences, \\
         \it \small
         Novosibirsk State University, \\
         \it \small
         Laurentyev 11, Novosibirsk, \\
         \it \small
         630090, Russia \\
         Presented by M.N.Achasov}

\date{}
\maketitle

\begin{abstract}
 The review of experimental results obtained with SND detector at
 VEPP-2M $e^+e^-$ collider in the energy region $\sqrt[]{s}=0.36$ -- $1.38$ GeV
 is given. The presented results include the following items: studies of the 
 light vector mesons radiative decays, OZI-rule and G-parity suppressed
 $\phi$-meson rare decays, $\phi$-meson parameters measurements, studies of 
 $e^+e^-\to\pi^+\pi^-\pi^0$ process dynamics, $\eta$ and $K_S$ mesons rare
 decays, $\eta$ and $\phi$ mesons conversion decays, and study of the $e^+e^-$
 annihilation into hadrons.
\end{abstract}

 The Spherical Neutral Detector (SND) operated since 1995 up to 2000 at
 VEPP-2M  \cite{vepp2m} $e^+e^-$ collider in the energy range from 0.36 to
 1.38 GeV. SND was described in detail in Ref.\cite{snd}.
 During six experimental years SND had collected data with integrated
 luminosity about 30 pb$^{-1}$. 

\begin{center}
\large \bf Radiative decays of the $\phi$, $\omega$, $\rho$ mesons
\end{center}

\begin{figure}
\includegraphics[width=0.5\textwidth]{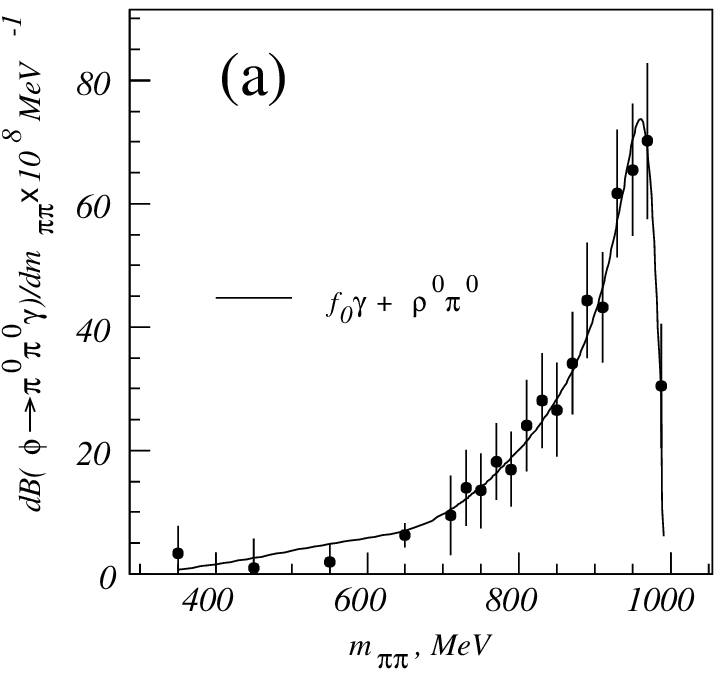}
\hfill
\includegraphics[width=0.5\textwidth]{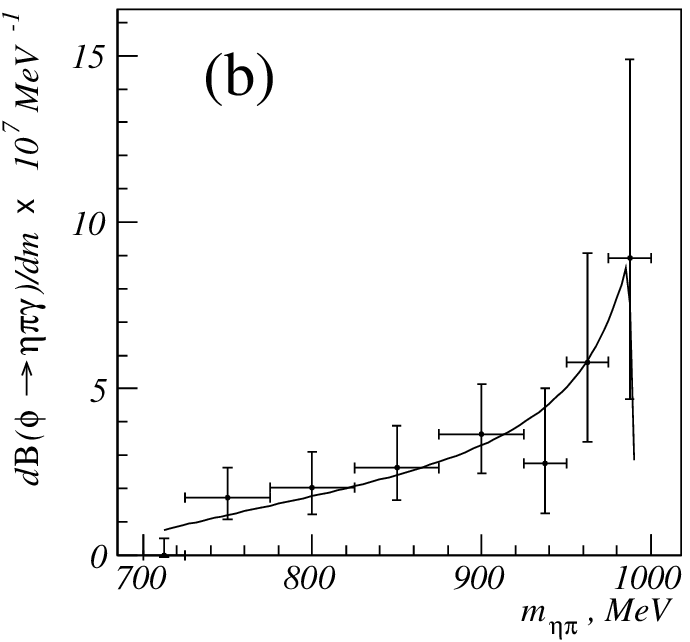}
\\
\caption{The $\pi^0\pi^0$ mass in the decay
 $\phi\to\pi^0\pi^0\gamma$~ (a) and the $\eta\pi^0$ mass in the decay
 $\phi\to\eta\pi^0\gamma$~ (b)}
\label{pipietapi}
\end{figure}

 {\bf Electric dipole transitions of the $\phi$ meson.}
 Till recently $\phi$ meson electric dipole transitions were not observed.
 A search for such decays was first performed with ND detector at VEPP-2M and
 the upper limits of about $10^{-3}$ were obtained \cite{nd1}. About the same
 time the theoretical proposal of the $\phi\to f_0\gamma$, $a_0\gamma$ decays
 search appeared \cite{annivn}. In 1997 the $\phi\to\pi^0\pi^0\gamma$,
 $\eta\pi^0\gamma$ decays were observed with SND
 \cite{phrad14}. The SND results based
 on the full data sample look as follows \cite{prad56}:
 $B(\phi\to\pi^0\pi^0\gamma) = (1.22 \pm 0.10 \pm 0.06) \cdot 10^{-4}$,
 $B(\phi\to\eta\pi^0\gamma)  = (0.88 \pm 0.14 \pm 0.09) \cdot 10^{-4}$.
 Studies of the $\pi^0\pi^0$ and $\eta\pi$ invariant mass spectra
 (Fig.\ref{pipietapi}) demonstrate that $f_0\gamma$ and $a_0\gamma$
 mechanisms dominate in these decays \cite{prad56}. So the following
 branching ratios were obtained:
 $B(\phi\to f_0\gamma) = (3.5 \pm 0.3 \pm^{1.3}_{0.5}) \cdot 10^{-4}$,
 $B(\phi\to a_0\gamma) = (0.88 \pm 0.14 \pm 0.09) \cdot 10^{-4}$.
 These relatively large values point out the exotic four-quark structure of
 $a_0$ and $f_0$ mesons \cite{anngub}. CMD2 measurements  reported in Ref.
 \cite{cmdrad12} agree with SND results. Also results of such measurements
 were recently reported by KLOE \cite{kloerad12}.
 
 {\bf The decays $\rho$, $\omega$ $\to$ $\pi^0\pi^0\gamma$.}
 In VDM model these decays proceed through the
 $\rho\to\omega\pi^0\to\pi^0\pi^0\gamma$ and
 $\omega\to\rho\pi^0\to\pi^0\pi^0\gamma$ transitions with the relative
 probability about $10^{-5}$ \cite{bramon1}. The same final state is also
 possible through the vector mesons radiative transitions to the $\pi^0\pi^0$
 scalar state with expected branching ratio
 about $1.4\cdot 10^{-5}$ \cite{marco,bramon2}.
 The only measurement of $\omega\to\pi^0\pi^0\gamma$ decay by GAMS \cite{gams}
 gives value of $(7.2\pm2.5) \cdot 10^{-5}$.
 The SND studies of these decays based on the one third of the accumulated
 statistics were already reported in Ref. \cite{roppg}.
 The results of a new analysis based on the full data sample of about
 9 pb$^{-1}$ are presented here.
\begin{figure}
\includegraphics[width=0.5\textwidth]{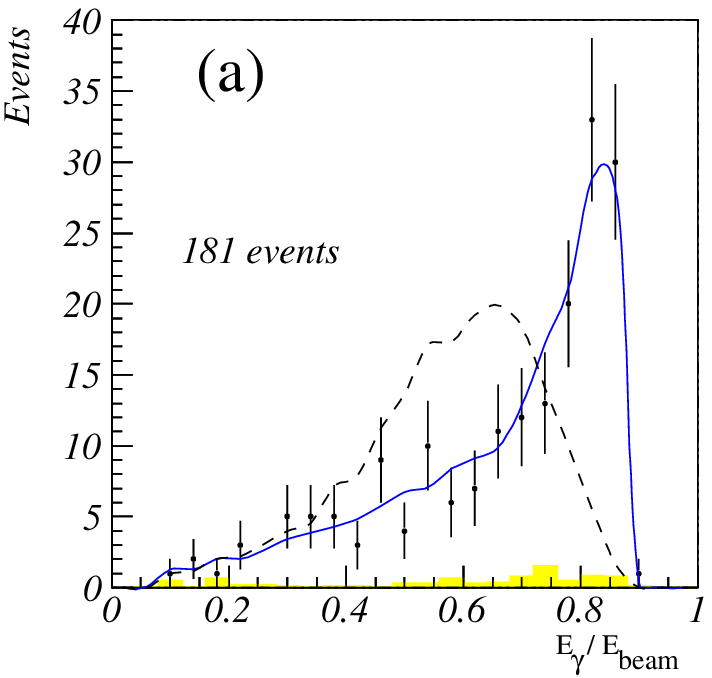}
\hfill
\includegraphics[width=0.5\textwidth]{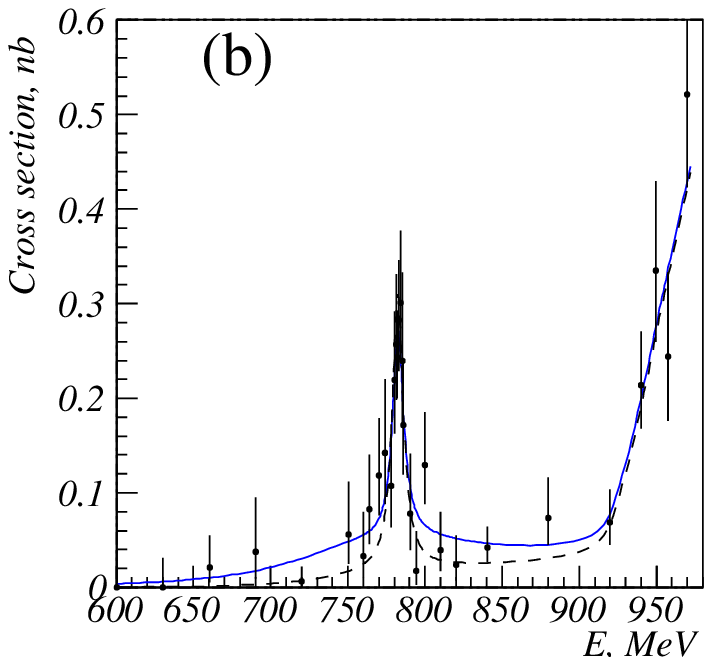}
\caption{(a) -- the photon energy spectrum in the reaction
         $e^+e^-\to\pi^0\pi^0\gamma$ in the energy range near $\omega$ meson
	 mass. Solid curve -- VDM model, dashed curve -- sum of VDM and
	 $\sigma\gamma$ mechanisms in case of destructive interference between
	 them. (b) -- cross-section for the $e^+e^- \to \pi^0\pi^0\gamma$
	 reaction. Solid line -- fit in VDM model, dashed line -- fit with sum
	 of VDM and $\rho\to S\gamma$ decay.}
\label{egfgcs00}
\end{figure}

 We cannot extract any information about $\omega$ decay mechanisms from the
 energy or angular distributions due to insufficient statistics. The photon
 energy spectrum shape agrees with VDM as well as with the sum of VDM and
 $V\to S\gamma$ ($V$ denotes vector meson and $S$ -- the scalar one, for
 example $\sigma$ meson) radiative decays mechanisms in case of constructive
 interference between them (Fig.\ref{egfgcs00} (a)). The destructive
 interference is ruled out experimental distribution. 

 The fit of the cross section (Fig.\ref{egfgcs00} (b)) included
 $\rho,\rho^\prime\to\omega\pi^0$ transitions and
 $\omega \to \pi^0\pi^0\gamma$ decay in VDM model and through
 $V\to S\gamma$ transitions. The strong difference in the energy dependences of
 the phase space for $\rho\to\omega\pi^0$ and $\rho\to S\gamma$ mechanisms
 allows to distinguish the different models. The model without
 $\rho\to S\gamma$ contribution gives $P(\chi^2)\simeq 1\%$. Inclusion of the
 scalar mechanism to the fit improves $P(\chi^2)$ to 30\%. The results of the
 fit follows:
 $B(\omega\to\pi^0\pi^0\gamma)=(6.3\pm1.4\pm0.8)\cdot10^{-5}$,
 $B(\rho\to\pi^0\pi^0\gamma)=(4.0\pm0.9\pm0.4)\cdot 10^{-5}$,
 $B(\rho\to S\gamma\to\pi^0\pi^0\gamma)=(2.0\pm0.7\pm0.3)\cdot10^{-5}$.
 So we confirm the value of $\omega$ decay obtained by GAMS. The $\rho$ meson
 decay to $\pi^0\pi^0\gamma$ was observed for the first time.

 {\bf The magnetic dipole transitions of the light vector mesons.}
 The magnetic dipole radiative decays are traditional objects in the light
 meson spectroscopy. Only one decay of this type $\phi\to\eta^{\prime}\gamma$
 was not observed till recently. This decay was observed with CMD2
 detector \cite{cmdrad3} and then confirmed by SND \cite{sndrad6}. The results
 of SND studies of the $\phi\to\eta^{\prime}\gamma$ in comparison with CMD2
 and KLOE measurements are listed in Table~\ref{tab1}.
\begin{table}[h]
\small
\hspace*{-0.2cm}
\begin{tabular}[h]{cccccc}
\hline
 & SND \cite{sndrad6} & SND & SND & CMD2 & KLOE  \\
 & $\eta^{\prime}\to\pi^+\pi^-\eta$  & $\eta^{\prime}\to\pi^0\pi^0\eta$
 & (average) & \cite{cmdrad4} & \cite{kloerad1} \\  \hline
 $B(\phi\to\eta^{\prime}\gamma)\cdot 10^5$ & $6.7\pm^{3.4}_{2.9}$ &
 $4.3\pm 1.6 \pm 0.9$ & $4.9 \pm^{1.6}_{1.5}$ & $6.4 \pm 1.6$ &
 $6.8 \pm 0.8$ \\ \hline
\end{tabular}
\caption{The comparison of the $B(\phi\to\eta^\prime\gamma)$ obtained with SND
         and results of the other experiments \cite{cmdrad4,kloerad1}}
\label{tab1}
\end{table}

\begin{figure}[t]
\includegraphics[width=0.5\textwidth,height=0.5\textwidth]{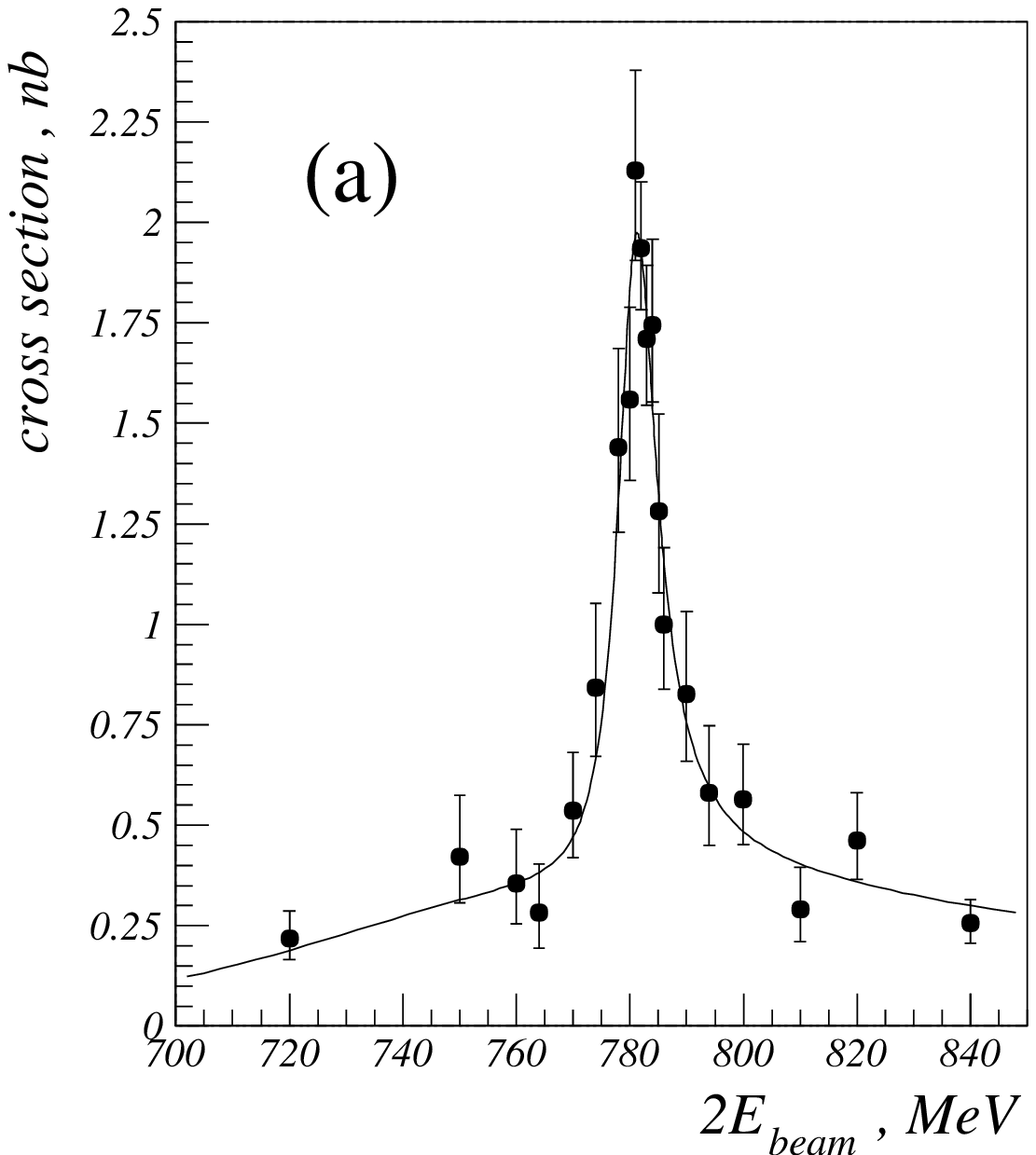}
\hfill
\includegraphics[width=0.5\textwidth,height=0.5\textwidth]{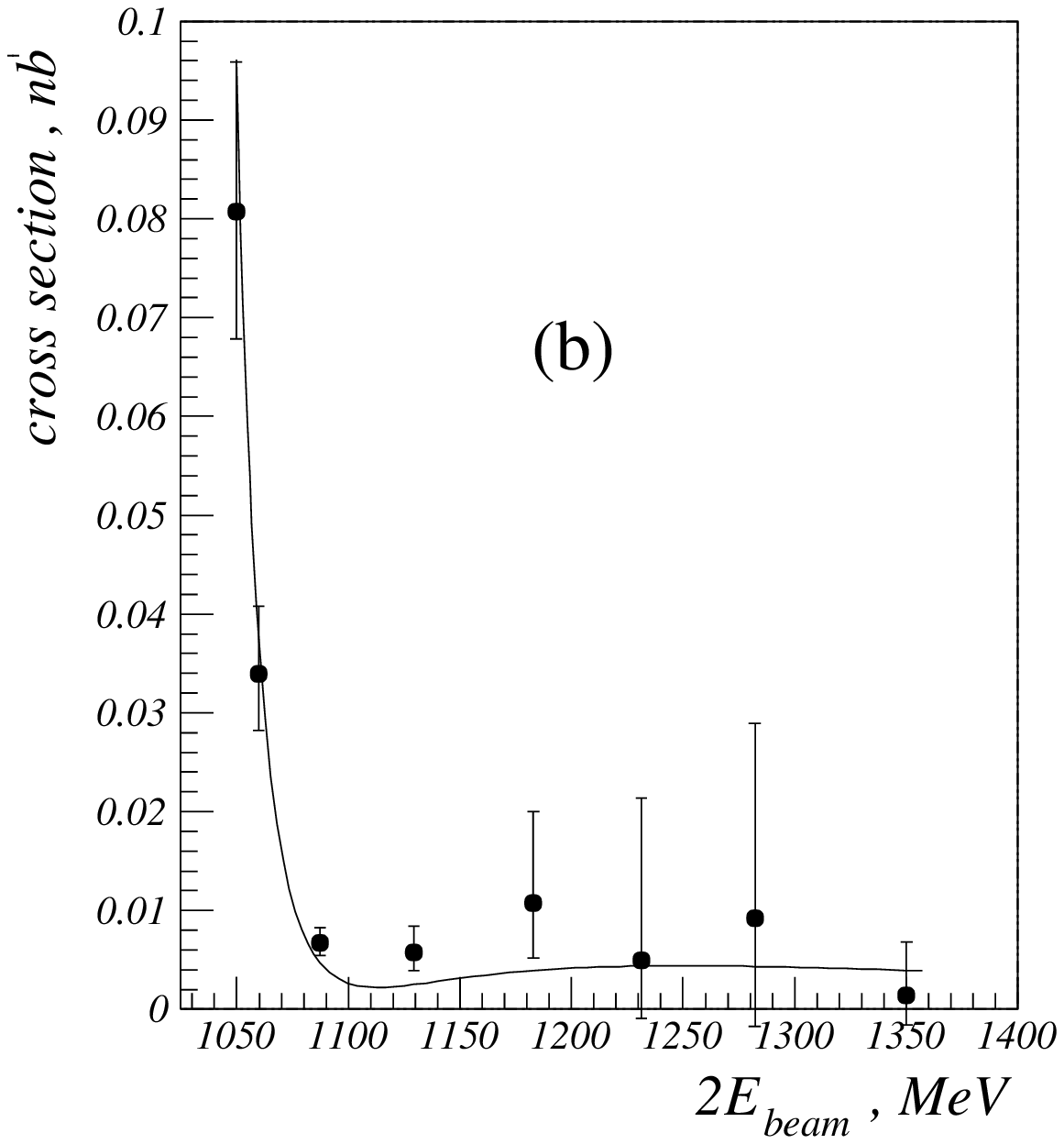}
\\
\caption{The cross section of the reaction $e^+e^-\to\eta\gamma$ in the $\rho$
         and $\omega$ energy region (a) and above $\phi$ meson (b)}
\label{cr_ome_phi}
\end{figure}

 The process $e^+e^-\to\eta\gamma$ in the seven photon final state was studied
 in full available energy region. The results based on the part
 of accumulated statistics were already published \cite{sndrad7}. Here we
 present the result obtained using full data set.
 It was found that cross section (Fig.\ref{cr_ome_phi}) can be described by
 sum of $\rho,\omega$ and $\phi$ resonances contributions only. The branching
 ratios obtained from the fit are presented in Table~\ref{tab2356}.
 The experimental ratio of the partial width
 $\Gamma_{\omega\eta\gamma}:\Gamma_{\rho\eta\gamma}:\Gamma_{\phi\eta\gamma}=1:(11.7 \pm 1.9):(15.9 \pm 1.9)$
 is consistent with a prediction of the simple quark model 1:8:12.

 The probability of the $\phi\to\eta\gamma$ decay was measured by SND in two
 other $\eta$ meson decay modes: $\eta\to\pi^+\pi^-\pi^0$ \cite{sndrad8} and
 $\eta\to\gamma\gamma$ \cite{sndrad9}. Combining the results of the three
 different modes the SND average was obtained: 
 $B(\phi\to\eta\gamma)=(1.310 \pm 0.045) \%$.~
 It is the most precise measurement of this value.

 The process $e^+e^-\to\pi^0\gamma$ was studied in the vicinity of
 $\phi$-meson \cite{sndrad9} and in the $\rho,\omega$ energy region \cite{radkorol}.
 As in previous case the cross section can be described by sum of
 $\rho,\omega$ and $\phi$ mesons only. The obtained branching ratios are
 listed in Table~\ref{tab2356}.
 The $\rho$ and $\omega$ branching ratios
 are in good agreement with both PDG values and prediction of a simple quark
 model. These results are based on a one third of available statistics. For
 full data sample we expect improvement of accuracy of the $\rho$ meson
 branching ratio. We also hope that combined analysis of data from $\phi$ and
 $\rho,\omega$ energy regions could reduce the systematic error of
 $\phi\to\pi^0\gamma$ branching ratio caused by the model dependence of
 $\phi-\omega$ interference description.

\begin{center} 
\large \bf $\phi$ meson energy region study
\end{center}

 {\bf OZI rule and G-parity suppressed $\phi\to \omega\pi^0$ and $\pi^+\pi^-$
 decays.}
 Till recently only one decay of this type $\phi\to\pi^+\pi^+$ was observed
 by detectors OLYA \cite{olya1} and ND \cite{nd3}. Such decays are possible
 through the $\omega-\phi$ mixing or direct transition \cite{theozi1,theozi2}.
 In the SND experiment the $\phi\to\pi^+\pi^+$ decay was studied \cite{sndpp}
 and the decay $\phi\to\omega\pi$ was observed for the first time
 \cite{sndwp13}.
 
 These decays are seen as interference patterns around $\phi$-resonance in the
 energy dependence of the cross section.
% (Fig.\ref{phipipiompi}).
 The Born cross section can be written as follows \cite{theozi1}:
 $$ \sigma(s)=\sigma_0(s)\times
  \biggl|1-Z{{m_\phi\Gamma_\phi}\over{D_\phi(s)}}\biggr|^2, $$
 where $\sigma_0$ is nonresonant cross section and $Z$ is complex interference
 amplitude. The measured branching ratios are listed in Table~\ref{tab2356}.
 The imaginary parts of $Z$ amplitudes: Im$(Z_{\pi\pi})=-0.041\pm 0.007$,
 Im$(Z_{\omega\pi})=-0.125\pm0.020$ agree with theoretical predictions based
 on standard $\omega-\phi$ mixing, while the expected values of the real parts
 exceed our results:Re$(Z_{\pi\pi})=0.061\pm 0.006$,
 Re$(Z_{\omega\pi})=0.108\pm 0.16$. The possible cause of this
 disagreement could be a nonstandard $\omega-\phi$ mixing or direct decays.

 {\bf $\phi$ meson parameters study.}
 The main parameters of the $\phi$ meson were measured through studies of the
 processes $e^+e^-\to K^+K^-, K_SK_L$ and  $\pi^+\pi^- \pi^0$ \cite{phi98}.
 The measured cross sections were approximated within the VDM, taking into
 account $\rho$, $\omega$ and $\phi$ mesons. Contributions from higher
 resonances $\rho^\prime$, $\omega^\prime$, $\phi^\prime$ were included
 in each cross section as constant terms. The $K^+K^-$ and $K_SK_L$ cross
 sections can be fitted by a sum of $\rho,\omega$ and $\phi$ contributions
 only, while for a good approximation of the $e^+e^- \to \pi^+\pi^-\pi^0$
 cross section the additional contribution, which can be attributed to the
 higher resonances, is strongly required. The obtained $\phi$-meson parameters
 (Table~\ref{tab2356}) mainly agree with PDG data and have accuracies
 comparable with the world averages. The only measured value which is in
 conflict with the now days world average is $\phi$-meson width.
 The world average $\Gamma_\phi$ value is strongly based on CMD2
 measurement $\Gamma_{\phi}=4.477\pm 0.036\pm 0.022$ MeV \cite{cmdgamma} which
 contradict to the SND one. But the recent CMD2 result
 $\Gamma_{\phi}=4.280\pm 0.033\pm 0.025$ MeV \cite{cmdgamma2}
 agreed with SND measurement.
 
 The $\phi$-meson leptonic branching ratio was also measured using
 $e^+e^-\to\mu^+\mu^-$ reaction \cite{sndmumu12}:
 $\sqrt[]{B(\phi\to e^+e^-)B(\phi\to\mu^+\mu^-)}=(2.93\pm0.11)\cdot 10^{-4}$,
 which is in a good agreement with branching value of $\phi\to e^+e^-$ decay.
 Using SND value of $\phi\to e^+e^-$ decay width we obtained the following
 leptonic branching ratio: $B(\phi \to l^+l^-)=(2.93\pm0.09)\cdot 10^{-4}$.
 
 {\bf The $e^+e^-\to\pi^+\pi^-\pi^0$ dynamics study and other results.}
 In SND experiment the dipion mass spectra were studied in the
 $e^+e^-\to\pi^+\pi^-\pi^0$ process in the energy region around $\phi$-meson
 \cite{sndspec}. Such studies provide the information about reaction dynamics
 as well as about $\rho$-meson parameters -- mass and width, $\rho^\pm$ and
 $\rho^0$ mass difference \cite{rhomass}. Spectra were analyzed within the VDM
 framework taking into account $\rho\pi$ transition, $\rho-\omega$ mixing and
 possible transition through intermediate states different from $\rho\pi$
 (for example, via $\rho^\prime\pi$). It was found that the experimental data
 can be described as a pure $\rho\pi$ transition. Upper limit on the branching
 ratio of the non $\rho\pi$ $\phi(1020)\to 3\pi$ decay was obtained:
 $B(\phi\to\pi^+\pi^-\pi^0) < 6 \cdot 10^{-4}$. This
 result agrees with CMD2 similar studies \cite{cmdspec}. Also the result of
 such studies was reported by KLOE \cite{kloespec}, but unfortunately the
 information given there is insufficient to do the comparison of the
\begin{table}
\caption{The results of the $\rho,\omega,\phi\to\eta\gamma$ decays
         studies using the seven photon final state, results of
	 $\rho,\omega,\phi\to\pi^0\gamma$, $\phi\to\pi^+\pi^-, \omega\pi$
	 decays measurements, the obtained $\phi$-meson parameters, results on
	 $\eta$ and $K_s$ mesons rare decays and conversion decays of $\eta$
	 and $\phi$ mesons}
\label{tab2356}
\small
\begin{tabular}{llll}
\hline
 & SND& Other data \\ \hline
 $B(\rho\to\eta\gamma)\cdot 10^4$&$2.77\pm 0.26\pm 0.16$&$3.28\pm0.37\pm0.23$ &(CMD2 \cite{cmdrad5}) \\
 $B(\omega\to\eta\gamma)\cdot 10^4$&$4.22\pm 0.47\pm 0.17$
 &$5.10\pm 0.72\pm 0.34$ &(CMD2 \cite{cmdrad5}) \\
 $B(\phi\to\eta\gamma)\cdot10^2$&$1.34\pm0.01\pm0.05$&$1.287\pm0.013\pm0.063$
  &(CMD2 \cite{cmdrad5})\\ \hline
 $B(\rho\to\pi^0\gamma)\cdot10^{4}$&$5.03\pm1.17\pm0.83$&$6.8\pm1.7$ &(PDG-2000) \\
 $B(\rho\to\pi^\pm\gamma)\cdot 10^{4}$&&$4.5\pm0.5$ &(PDG-2000)\\ 
 $B(\omega\to\pi^0\gamma)\cdot10^{2}$&$9.17\pm0.16\pm0.46$&$8.5\pm0.5$ &(PDG-2000)\\
 $B(\phi\to\pi^0\gamma)\cdot10^{3}$&$1.23\pm0.04\pm0.09$&$1.26\pm0.10$ &(PDG-2000)\\ \hline
 $B(\phi\to\pi^+\pi^-)\cdot 10^{5}$&$7.1\pm 1.4$&$8\pm^5_4$&\cite{olya1,nd3} \\
 $B(\phi\to\omega\pi^0)\cdot 10^{5}$&$5.2\pm^{1.3}_{1.1}$&& \\ \hline
 $m_\phi$, MeV&$1019.42\pm0.02 \pm 0.04$&$1019.417\pm0.014$ &(PDG-2000) \\ 
 $\Gamma_\phi$, MeV&$4.21\pm 0.03\pm 0.02$&$4.458 \pm 0.032$ &(PDG-2000) \\
 $B(\phi\to e^+e^-)\cdot 10^4$&$2.93\pm 0.02\pm 0.14$&$2.91 \pm 0.07$ &(PDG-2000) \\
 $B(\phi\to K^+K^-)$, \%&$47.6\pm 0.3\pm 1.6$&$49.2 \pm 0.7$ &(PDG-2000) \\
 $B(\phi\to K_SK_L)$, \%&$35.1\pm 0.2\pm 1.2$&$33.8 \pm 0.6$ &(PDG-2000) \\
 $B(\phi\to \pi^+\pi^-\pi^0)$, \%&$15.9\pm 0.2\pm 0.8$&$15.5 \pm 0.6$ &(PDG-2000) \\
 $B(\phi\to \eta\gamma)$, \%&$1.33\pm 0.03\pm 0.05$&$1.297 \pm 0.033$ &(PDG-2000) \\ 
\hline
 $B(\eta\to\pi^0\pi^0)\cdot10^4$&$<6$ \cite{sndeta1}&$<4.3$ &(CMD2 \cite{cmdrad12}) \\
 $B(\eta\to\pi^0\gamma\gamma)\cdot10^4$&$<8.4$ \cite{sndeta2}&$7.1\pm1.4$ &(PDG2000) \\
 $B(K_S\to3\pi^0)\cdot10^5$&$<1.4$ \cite{sndks}&$<1.9$ &(CPLEAR \cite{cpks}) \\
 $B(\phi\to\eta e^+e^-)\cdot10^4$&$1.19\pm0.22$ \cite{sndcon}&$1.17\pm0.12$ &(CMD2 \cite{cmdcon1}) \\
 $B(\eta\to e^+e^-\gamma)\cdot10^3$&$5.15\pm0.96$ \cite{sndcon}&$7.10\pm0.79$ &(CMD2 \cite{cmdcon1}) \\
 $B(\phi\to\pi^0e^+e^-)\cdot10^5$&$1.05\pm0.37$&$1.22\pm0.40$ &(CMD2 \cite{cmdcon2}) \\ \hline
\end{tabular}
\end{table}
 results. Neutral and charged $\rho$-mesons mass difference was found to be
 consistent with zero: $m_{\rho^\pm}-m_{\rho^0}=-1.3\pm2.3$ MeV. The
 $\rho$-meson mass and width were measured equal to $m_{\rho}=775.0\pm1.3$ MeV,
 $\Gamma_{\rho}=150.4\pm3.0$ MeV.
 The $\rho$ mass values obtained by using different reactions contradict
 each other. SND $\rho$-mass value support the results of the $e^+e^-$
 annihilation and $\tau$ decay experiments:$m_\rho=776\pm 0.9$ MeV. But the
 PDG value $769.3\pm 0.8$ MeV, which takes into account all experiments in
 which the $\rho$-meson mass was measured, contradicts our result.

 Some other results obtained using statistics collected in the $\phi$-meson
 energy region are presented in Table~\ref{tab2356}.

\begin{center}
\large \bf $e^+e^-$ annihilation into hadrons above 1 GeV
\end{center}

 The light vector mesons are studied rather well. They are 2 quark states,
 their masses, widths and the main decays are measured with high accuracy.
 The experimental data also point out the existence of the states with vector
 meson quantum numbers $I^G(J^{PC}) = 0^+(1^{--}), 0^-(1^{--})$ and
 masses above 1 GeV. Parameters of these states are not well established due
 to the poor accuracy and conflicting of experimental data. The nature of
 these states is not clear. They are considered as a mixture of two quark,
 four quark
 and hybrids states \cite{don} or as a two quark states -- radial and orbital
 excitations of the $\rho$, $\omega$ and $\phi$ mesons \cite{ak}. In this
 context the main experimental task is the improvement of the cross sections
 measurement accuracy. In SND experiment the following processes were studied.

\begin{figure}
\includegraphics[width=0.5\textwidth,height=0.4\textwidth]{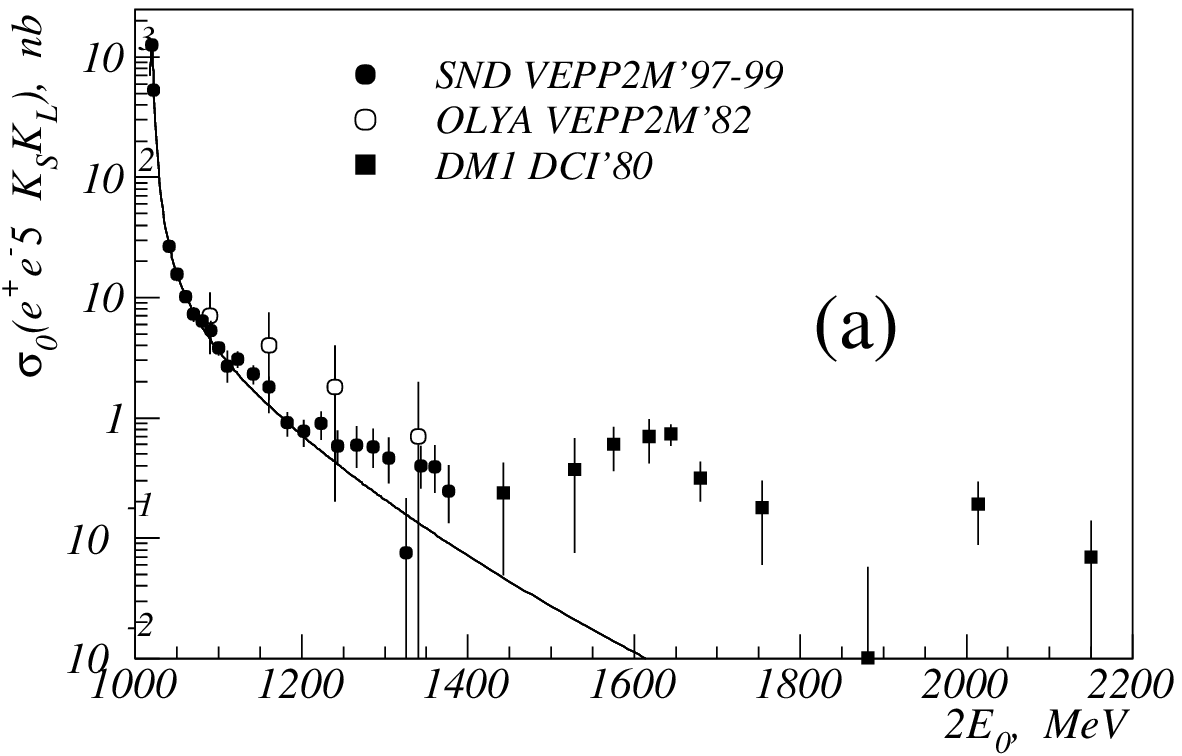}
\hfill
\includegraphics[width=0.5\textwidth,height=0.4\textwidth]{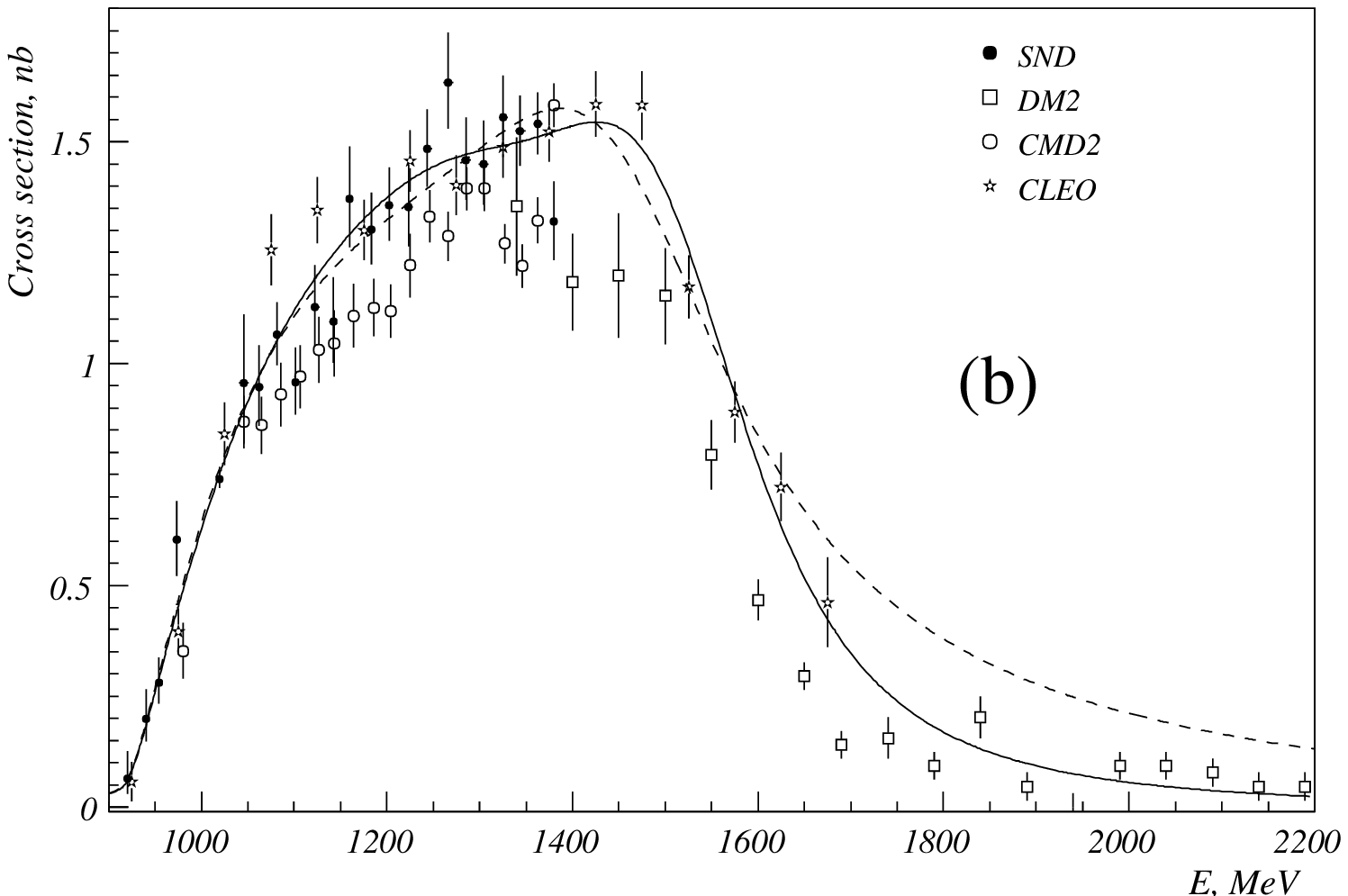}
\caption{(a) -- the cross section of the reaction $e^+e^- \to K_SK_L$. The
         results of the SND, OLYA \cite{olya2} and DM1 \cite{dm1} are shown.
	 Curve is theoretical cross section in conventional VDM.
	 (b) -- the cross section of the reaction
	 $e^+e^- \to \omega\pi \to \pi^0 \pi^0 \gamma$. The results of the
	 SND \cite{sndompn}, DM2 \cite{dm21}, CMD2 \cite{cmdmhad1} and
	 CLEO2 \cite{cleo} are shown. Curves are results of fitting in models
	 described in Ref.\cite{sndompn}}
\label{crsecppg}
\end{figure}

 The $e^+e^- \to K_SK_L$ cross section was measured using $K_S \to \pi^0\pi^0$
 decay mode. Our measurements in comparison with OLYA and DM1 results are
 shown in Fig.\ref{crsecppg} (a). The curve is theoretical cross
 section with $\rho$, $\omega$ and $\phi$ contributions only. Experimental data
 above 1.2 GeV exceed the conventional VDM prediction.

 The process $e^+e^- \to \omega\pi$ was studied in the $\pi^0\pi^0\gamma$ final
 state \cite{sndompn}. Measured cross section in comparison with the other
 results is shown in Fig.~\ref{crsecppg} (b). The systematic error of SND
 measurement is about 5\%. The CLEO2 results\footnote{$e^+e^-$ annihilation
 cross section was calculated from $\tau\to 3\pi\pi^0$ decay data using CVS
 hypothesis} are in good agreement with ours, while CMD2 measurements are
 about 10\% lower, but this difference is smaller than the 15\% systematic
 error quoted in Ref.\cite{cmdmhad1}. The same process was also studied in the
 $\pi^+\pi^-\pi^0\pi^0$ final state \cite{sndompc}. Obtained cross section
 agrees with our result in $\pi^0\pi^0\gamma$ mode. Its systematic error was
 estimated to be 20\% for $\sqrt[]{s}<1150$ MeV and 15\% at $\sqrt[]{s}>1150$.

\begin{figure}
\includegraphics[width=0.45\textwidth]{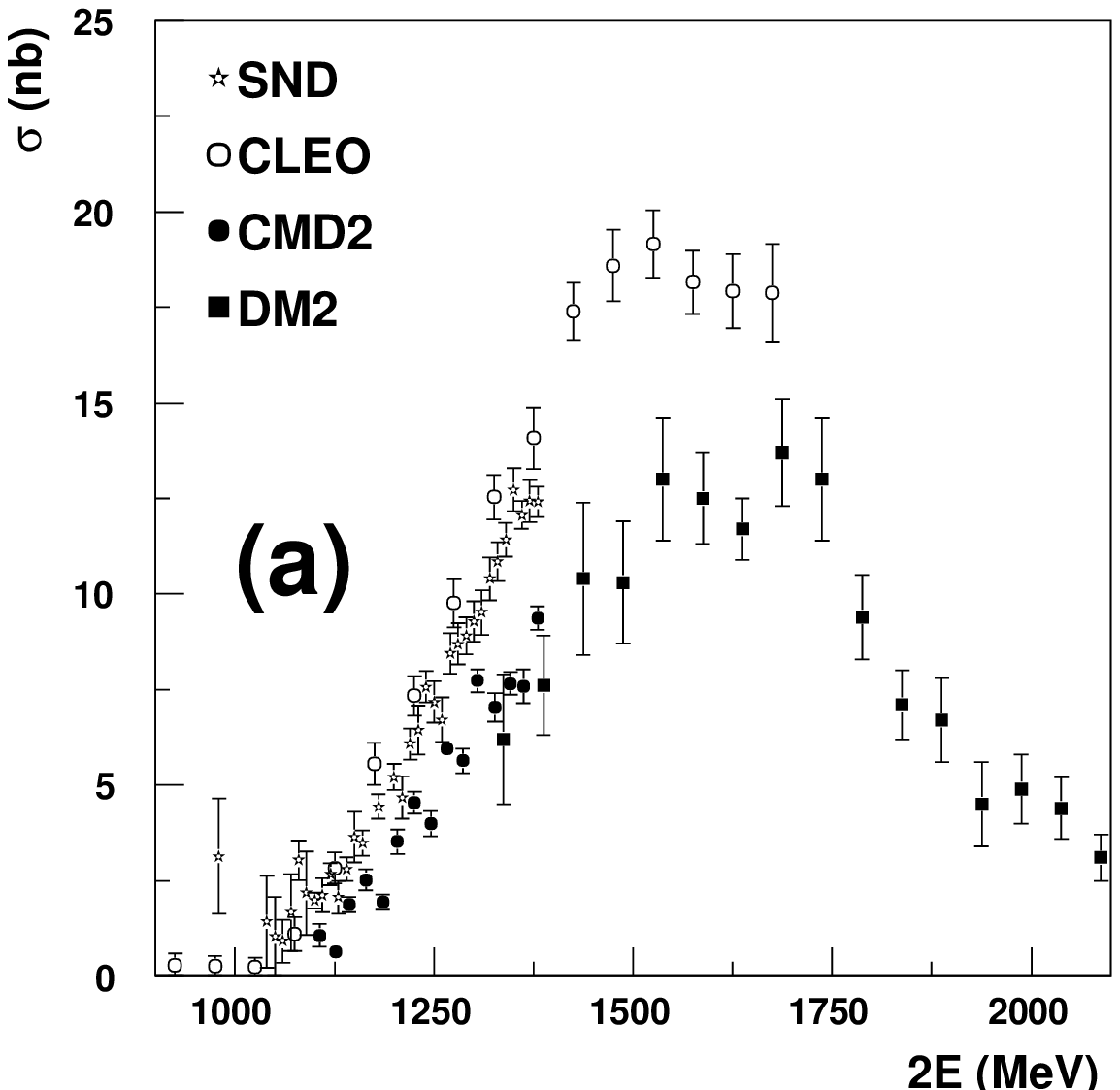}
\includegraphics[width=0.45\textwidth]{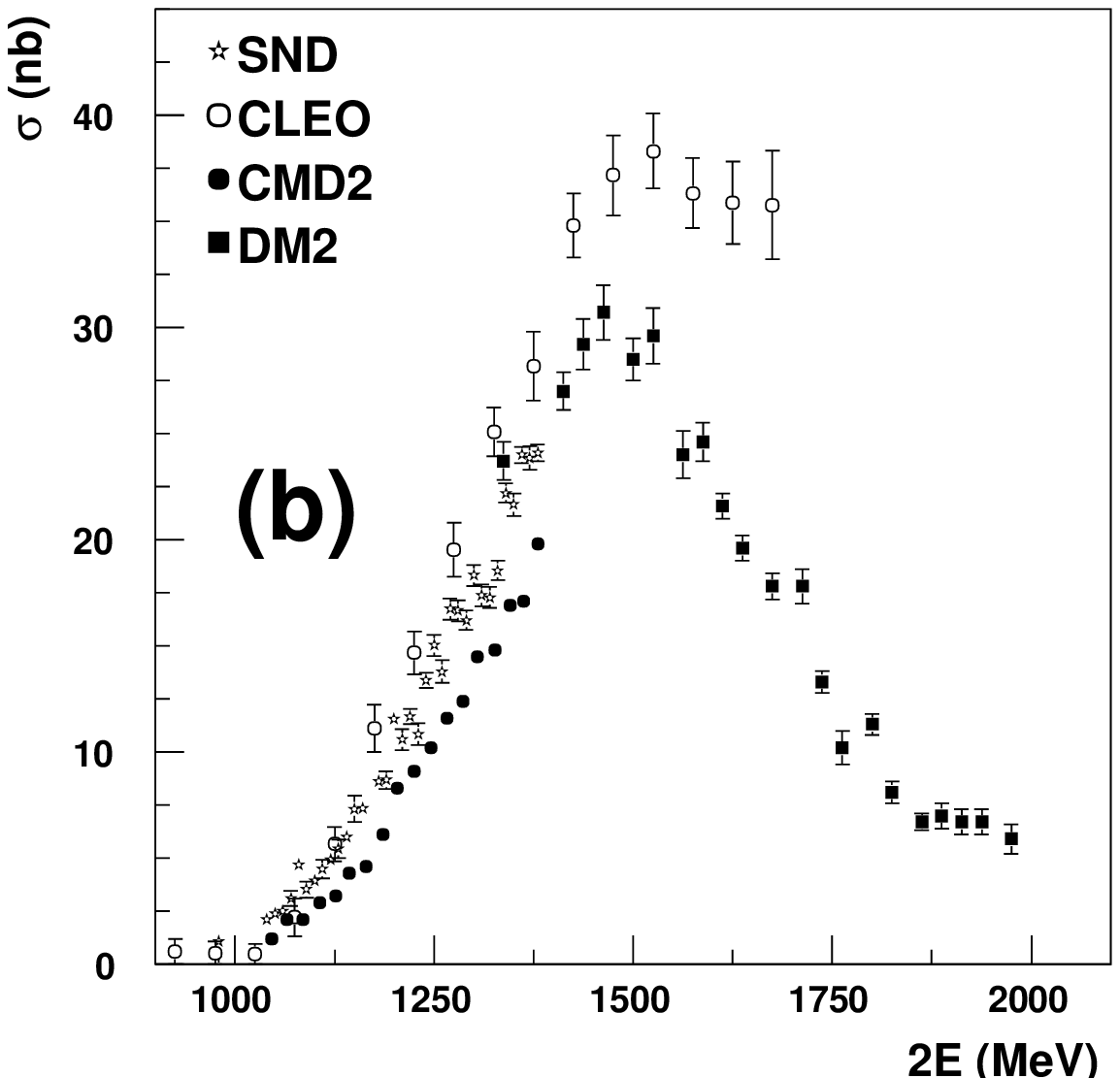}
\caption{(a) -- the cross section of the reaction
         $e^+e^- \to \pi^+\pi^-\pi^0\pi^0$ with subtracted contribution from
	 $\omega\pi^0$. 
         (b) -- the cross section of the reaction
         $e^+e^- \to \pi^+\pi^-\pi^+\pi^-$. The results of the 
	 SND \cite{sndompc}, DM2 \cite{dm22}, CMD2 \cite{cmdmhad1} and
	 CLEO2 \cite{cleo} are shown.}
\label{c4pnc}
\end{figure}

 The $e^+e^-\to\pi^+\pi^-\pi^0\pi^0$ cross section with subtracted contribution
 from $\omega\pi^0$ is shown in Fig.\ref{c4pnc} (a). The systematic error of
 SND measurements is about 20\%. The SND result is compared with CLEO2, CMD2
 and DM2 data\footnote{While extracting cross section data from CLEO2 results
 the ratio
 $\sigma_{e^+e^-\to\pi^+\pi^-\pi^+\pi^-}/\sigma_{e^+e^-\to\pi^+\pi^-2\pi^0}=2$
 was assumed, which is confirmed by SND measurements \cite{sndompc}}.
 The two groups are seen: CMD2 dots better agree with DM2 while
 SND ones -- with CLEO2. But the difference is smaller than systematic errors.
 The measured cross section of the $e^+e^-$ annihilation into four charged
 pions is shown in Fig.\ref{c4pnc} (b). Here SND dots better agree with DM2
 result. The systematic error of the $e^+e^-\to\pi^+\pi^-\pi^+\pi^-$  cross
 section was estimated to be 12\%  for $\sqrt[]{s}<1150$ MeV and  8\%
 at $\sqrt[]{s}>1150$.

\begin{figure}[t]
\includegraphics[width=0.5\textwidth,height=0.4\textwidth]{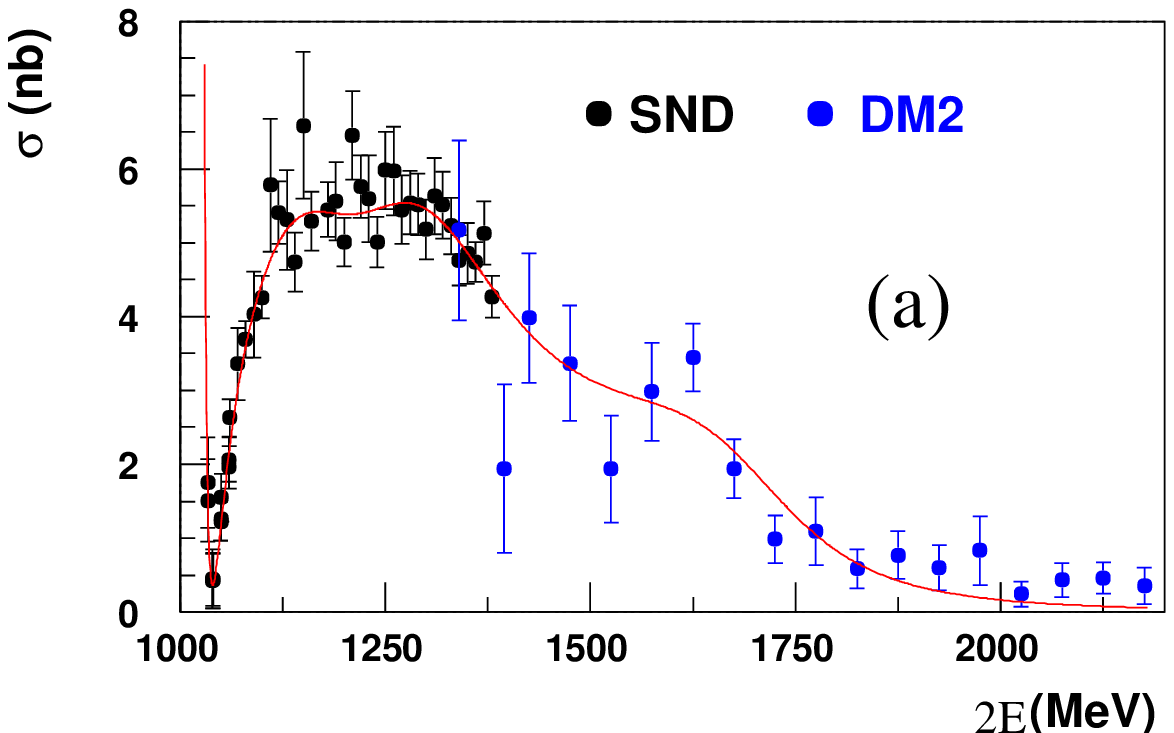}  
\hfill
\includegraphics[width=0.5\textwidth,height=0.4\textwidth]{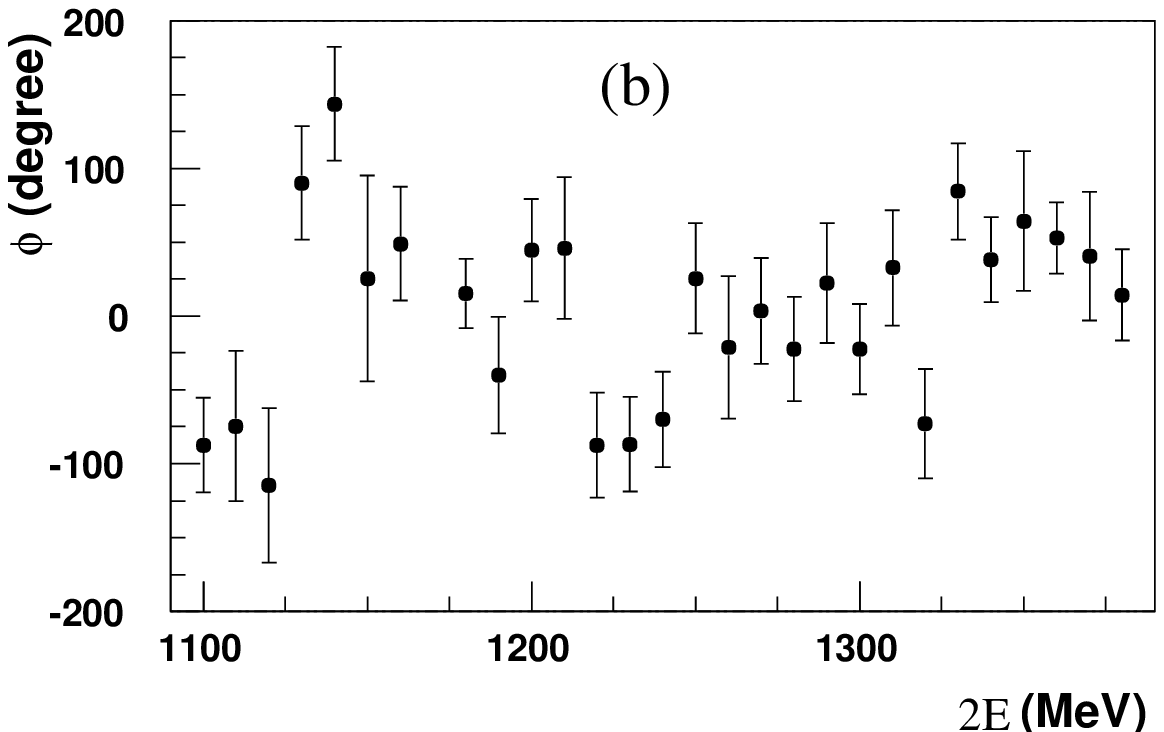}  
\caption{(a) -- the cross sections of the reactions
         $e^+e^- \to \pi^+\pi^-\pi^0$. The results of the SND and 
	  DM2 \cite{dm23} are shown. Curve is the fit by a sum of $\omega$,
	  $\phi$ and three $\omega^i$ amplitudes. (b) -- the energy
	  dependence of the relative phase between $A_{\rho\pi}$ and
	  $A_{\omega\pi}$ amplitudes.}
\label{pi3fitfaz}
\end{figure}

 Both $e^+e^-\to\rho\pi$ and $e^+e^-\to\omega\pi^0$ mechanisms contribute 
 to the $\pi^+\pi^-\pi^0$ final state. The $\omega\pi$ contribution
 was predicted in Ref.\cite{aomp} and observed by SND \cite{sndomp}.
 SND already reported the total cross section measurements based on the part
 of the accumulated statistics \cite{sndpi3mhad}. The new result of the cross
 section measurement is presented here. For data analysis we use the following
 theoretical model, taking into account the $\rho-\omega$ mixing \cite{aomp}:
 $$
 {d\sigma \over dm_0 dm_+} = { {4\pi\alpha} \over {s^{3/2}} }
 {{|\vec{p}_+ \times \vec{p}_-|^2} \over {12\pi^2\mbox{~}\sqrt[]{s}}} m_0m_+
 \cdot \biggl|A_{\rho\pi}(s) \sum_{i=+,0,-}
 { g_{\rho^i\pi\pi} \over D_\rho(m_i)} +
 A_{\omega\pi}(s)
 {\Pi_{\rho\omega}g_{\rho^0\pi\pi}\over D_\rho(m_0) D_\omega(m_0)}\biggr|^2
 $$
 $$
 A_{\rho\pi} \sim \sum_{V=\omega,\phi,\omega^\prime,{\ldots} }
 {{\Gamma_Vm_V^2\sqrt[]{m_V\sigma(V\to\rho\pi)}} \over {D_V(s)} }
 \mbox{~~~~~~} A_{\omega\pi}(s) = \sum_{V=\rho,\rho^\prime,{\ldots} }
  {g_{\gamma V}g_{V\omega\pi^0} \over D_V(s)}
  $$
  $$
  \mbox{Im}(\Pi_{\rho\omega}) \ll \mbox{Re}(\Pi_{\rho\omega}), \mbox{~~~~}
  \mbox{Re}(\Pi_{\rho\omega}) = 2m_\omega\delta, \mbox{~~~~} \delta = 2.3
  \mbox{~MeV},
  \mbox{~~~~} \delta \sim \sqrt[]{B(\omega\to\pi^+\pi^-)}
  $$ 
  The combined studies of the total cross section and dipion mass spectra
  provide the information about relative phase between $A_{\rho\pi}$ and
  $A_{\omega\pi}$ amplitudes and $\omega\to\pi^+\pi^-$ branching ratio.
  We performed the combined fit of $\pi^+\pi^-\pi^0$ (Fig.\ref{pi3fitfaz}) and
  $\omega\pi^+\pi^-$ cross sections. The cross section measured by SND in
  the $\phi$-meson energy region was also included in the fit. The best
  description of the data was obtained when the cross sections were fitted by
  a sum of $\omega$, $\phi$ and three $\omega^i$ amplitudes. The obtained
  $\omega^i$ parameters are listed in Table~\ref{tab7}.
\begin{table}
\begin{center}
\caption{$\omega^i$ parameters obtained from the fit. Here $\phi$ denotes
a relative phases between $\omega$ and $\omega^i$ primes.}
\label{tab7}
\small
\begin{tabular}{llll}
\hline
 &$\omega^{1}$&$\omega^{2}$&$\omega^{3}$ \\ \hline
$m$, MeV&1250$\pm$29&1400$\pm$19&1771$\pm$28 \\
$\Gamma$, MeV&426$\pm$135&626$\pm$89&473$\pm$76 \\
$\sigma(V\to\rho\pi)$, nb&0.56$\pm$0.25&3.90$\pm$0.39&2.28$\pm$0.46 \\
$\sigma(V\to\omega\pi\pi)$, nb&0&0.046$\pm$0.039&2.49$\pm$0.33 \\
$\phi$&$\pi$&$\pi$&0 \\
%$B(V\to 3\pi)$&$1$&$>99$&$\sim 49$ \\
%$B(V\to\omega\pi\pi)$&0&$<0.01$&$\sim 0.51$ \\
$\Gamma(V\to e^+e^-)$, eV&$\sim 25$&$\sim 300$&$\sim 470$ \\ \hline
\end{tabular}
\end{center}
\end{table}

 To obtain relative phase between $A_{\rho\pi}$ and $A_{\omega\pi}$ amplitudes
 and $\omega\to\pi^+\pi^-$ branching ratio the invariant mass distribution and
 the ratio of $\rho\pi$ to $3\pi$ cross sections were fitted together in each
 energy point. The energy dependence of the relative phase is shown in
 Fig.\ref{pi3fitfaz}. The obtained branching ratio
 $B(\omega\to\pi^+\pi^-)=2.46\pm0.42\pm0.15$ agrees with world average value.

 A search for direct production of $a_2$ and $f_2$ mesons in $e^+e^-$
 annihilation was performed with SND \cite{snda2f2}. The following upper 
 limits were obtained $\Gamma(a_2\to e^+e^-)<0.56$ eV and
 $\Gamma(f_2\to e^+e^-)<0.11$ eV. These upper limits are only
 four times higher than unitarity limit \cite{a2f2the}.

\begin{center} 
\large \bf Conclusion
\end{center}

 The SND detector operated since 1995 up to 2000 at VEPP-2M collider in
 the energy range $360 < \sqrt[]{s} < 1380$ MeV and had collected data
 with integrated luminosity of about $30$ pb$^{-1}$. The $\rho$, $\omega$,
 $\phi$ mesons decays and $e^+e^-$ annihilation into hadrons were studied.
 New rare decays $\phi\to\pi^0\pi^0\gamma$, $\eta\pi^0\gamma$,
 $\omega\pi^0$ and $\rho\to\pi^0\pi^0\gamma$ were observed. Many other results
 were obtained.

 The present work was supported in part by grants RFBR 00-15-96802,
 01-02-22003, 01-02-16934-a, 00-02-17478, 00-02-17481,
 grant no. 78 1999 of Russian Academy of
 Science for young scientists.

\end{document}